\documentclass[aps,pre,reprint,superscriptaddress,twocolumn,longbibliography,floatfix]{revtex4-2}

\usepackage{empheq}
\usepackage{graphicx}
\usepackage{dcolumn}
\usepackage{bm}

\usepackage[utf8]{inputenc}
\usepackage[T1]{fontenc}
\usepackage{mathptmx}
\usepackage{etoolbox}
\usepackage{xcolor}
\usepackage{mathpazo}
\usepackage{bm}
\usepackage[unicode=true,
 bookmarks=true,bookmarksnumbered=false,bookmarksopen=false,
 breaklinks=false,pdfborder={0 0 0},pdfborderstyle={},backref=false,colorlinks=true]
 {hyperref}
\hypersetup{pdfborderstyle={},pdfborderstyle={},pdfborderstyle={},pdfborderstyle={},pdfborderstyle={},pdfborderstyle={},linkcolor=blue,citecolor=blue}
\usepackage{breakurl}

\usepackage{natbib}
\usepackage{amssymb,amsmath}

\newcommand{\bp}{\beta p_{\|}}
\newcommand{\gi}{\mathrm{g}}
\newcommand{\ii}{{\alpha}}
\newcommand{\lcp}{\lambda_{\text{cp}}}
\newcommand{\ex}{\text{ex}}

\newcommand{\beq}{\begin{equation}}
\newcommand{\eeq}{\end{equation}}
\newcommand{\beqa}{\begin{eqnarray}}
\newcommand{\eeqa}{\end{eqnarray}}
\newcommand{\nn}{\nonumber\\}
\newcommand{\xx}{{\|}}
\newcommand{\yy}{{\perp}}
\def\bal#1\eal{\begin{align}#1\end{align}}

\makeatletter
\def\@email#1#2{%
 \endgroup
 \patchcmd{\titleblock@produce}
  {\frontmatter@RRAPformat}
  {\frontmatter@RRAPformat{\produce@RRAP{*#1\href{mailto:#2}{#2}}}\frontmatter@RRAPformat}
  {}{}
}%
\makeatother
\begin{document}

\title[]{Exploring anisotropic pressure and spatial correlations in strongly confined hard-disk fluids. Exact results}
\author{Ana M. Montero}
\affiliation{Departamento de F\'isica, Universidad de Extremadura, E-06006 Badajoz, Spain}
\author{Andr\'es Santos}%
\affiliation{Departamento de F\'isica, Universidad de Extremadura, E-06006 Badajoz, Spain}
\affiliation{Instituto de Computaci\'on Cient\'ifica Avanzada (ICCAEx), Universidad de Extremadura, E-06006 Badajoz, Spain}

\date{\today}

\begin{abstract}
This study examines the transverse and longitudinal properties of hard disks confined in narrow channels. Employing an exact mapping of the system onto a one-dimensional polydisperse, nonadditive mixture of hard rods with equal chemical potentials, we compute various thermodynamic properties, including the transverse and longitudinal equations of state, along with their behaviors at both low and high densities. Structural properties are analyzed using the two-body correlation function and the radial distribution function, tailored for the highly anisotropic geometry of this system. The results are corroborated by computer simulations.
\end{abstract}

\maketitle

\paragraph*{Introduction.}
The investigation of fluids under extreme confinement has garnered considerable attention over the years, playing a pivotal role in comprehensively understanding liquid behavior. Among the various confined geometries in which liquids can be situated, quasi-one-dimensional (q1D) channels hold particular significance. In these configurations, the available space along one dimension (the longitudinal axis) vastly exceeds that along the perpendicular, confined axes. This disparity in dimensions characterizes the highly anisotropic nature of q1D confinement.
Thus, these q1D systems lie halfway between purely one-dimensional (1D) systems, which are known to have analytical solutions under certain circumstances~\cite{KT68,BB83a,BE02,HC04,F16,MS19,MS20}, and bulk two- or three-dimensional systems, whose properties are generally addressed through approximations, numerical solutions, or simulations~\cite{MBS97,LS04,KLM04,LKM05}.

In addition to their inherent theoretical interest, these systems have gained even greater relevance with the advancement of nanofluidics~\cite{CGI05}, nanopores~\cite{KHD11,LKK14,M24}, and various experimental techniques capable of replicating such conditions~\cite{CLSR02,CPR06,LVMCHR09,LZHRL17}. These experimental setups have provided invaluable insights into the behavior of fluids under extreme confinement, further motivating theoretical investigations into the properties of fluids in q1D channels.

The task of deriving exact, analytical expressions for the thermodynamic and structural properties of q1D systems has been a focal point of research over the years and has been approached from various theoretical perspectives and simulation methods~\cite{KP93,VBG11,GV13,GM14,HFC18,ZGM20,HBPT20,*HC21,*TPHB21,HC21,F23,FS24}. Exact results for the longitudinal thermodynamic properties of these systems are known, and more recently, exact results for their structural properties have also been obtained, although numerical integration is ultimately required~\cite{MS23,MS23b,MS24}. Purely analytical expressions found in the literature are typically obtained through approximations~\cite{KMP04,GM14,MS23,M14b}. Despite some advances in understanding transverse properties (see especially Refs.~\cite{VBG11,FS24}), a comprehensive study in this area is still lacking, and a unified methodology for investigating these systems remains elusive.

In this article, we investigate a q1D confined system characterized by one longitudinal dimension of length $L_\xx=L$ and one transverse dimension of length $L_\yy=\epsilon\ll L$. The particles in the system interact via a hardcore pairwise additive potential, with each particle having a hardcore diameter of $d=1$ (henceforth defining the unit of length), so that the separation between the two confining walls is  $1+\epsilon$ \footnote{Note that we have defined the transverse dimension $L_\yy$ not as the wall separation, but as the range of transverse positions accessible to the centers of the disks.}. The smallness of the transverse dimension prevents particles from bypassing each other, compelling them to arrange in a single-file formation along the longitudinal dimension. Moreover, we impose $\epsilon\leq\frac{\sqrt{3}}{2}$  to ensure that interactions with second-nearest neighbors are absent.

In these circumstances, it can be demonstrated that the confined q1D system is formally equivalent to a 1D polydisperse  mixture with equal chemical potential~\cite{MS23,MS23b,MS24}. Particles in the mixture are categorized into different species based on the transverse coordinates $y$ (with $-\epsilon/2\leq y \leq \epsilon/2$) of the disks  in the original system. They interact via an effective hardcore distance of $a_{y_1y_2}=\sqrt{1-y_{12}^2}$, where $y_{12}^2=(y_1-y_2)^2$ \footnote{Throughout the paper, subscripts denote dependence on the continuous variable $y$, highlighting the connection between the q1D fluid and the 1D polydisperse system.}. Since $a_{y_1y_2}\neq\frac{1}{2}\left(a_{y_1y_1}+a_{y_2y_2}\right)$, the 1D mixture is indeed a nonadditive one. The mole fraction distribution function, $\phi_y^2$, of the 1D polydisperse system coincides with the transverse density profile of the equivalent hard-disk confined fluid.

\paragraph*{The 1D polydisperse system.}
Typically, the exact solution for 1D fluids is derived within the isothermal-isobaric ensemble~\cite{S16}. In particular, the nearest-neighbor probability distribution function of a generic 1D polydisperse hard-rod fluid is  $P_{y_1y_2}^{(1)}(x)=(\phi_{y_2}/\phi_{y_1}) A_{y_1} A_{y_2}e^{-\bp x}\Theta (x-a_{y_1y_2})$, where $\Theta(\cdot)$ is the Heaviside step function, $\beta \equiv 1/k_\mathrm{B}T$ ($k_\mathrm{B}$ and $T$ being the Boltzmann constant and the absolute temperature, respectively), and $p_\xx$ is the 1D pressure, which has dimensions of  force. Given an {arbitrary} mole fraction distribution $\phi^2_y$, the function $A_y$ is the solution to~\cite{S16,MS23b}
\begin{equation}
\label{eq:eigen}
A_{y_1} \int_{\epsilon} {d}y_2\,e^{-\bp a_{y_1y_2}}A_{y_2}\phi_{y_2}=\bp {\phi_{y_1}}.
\end{equation}
Successive convolutions of $P_{y_1y_2}^{(1)}(x)$ yield the pair correlation function $g_{y_1y_2}(x)$. Its Laplace transform, $G_{y_1y_2}(s)=\int_0^\infty {d}x\, e^{-s x}g_{y_1y_2}(x)$, follows the integral equation~\cite{MS23b,MS24}
\bal
\label{Fredh}
\frac{\phi_{y_2}}{ A_{y_2}}G_{y_1 y_2}(s)=&\int_{\epsilon} {d}y_3\,\phi_{y_3}G_{y_1 y_3}(s)A_{y_3}\frac{e^{-(s+\bp)a_{y_2 y_3}}}{s+\bp}
\nn
&+\frac{A_{y_1}}{\lambda\phi_{y_1}}\frac{e^{-(s+\bp)a_{y_1 y_2}}}{s+\bp}.
\eal
Here, the linear density \footnote{The linear density $\lambda$ should be distinguished from the number of particles per unit area (areal density) $\rho\equiv N/L\epsilon=\lambda/\epsilon$.} $\lambda=N/L$ (where $N$ is the number of particles) is given by \cite{MS23,MS23b}
\beq
\label{eq:lambda}
\frac{\bp}{\lambda}=1+\int_{\epsilon} {d}y_1\int_{\epsilon} {d}y_2\,\phi_{y_1}\phi_{y_2}A_{y_1} A_{y_2}a_{y_1y_2} e^{-\bp a_{y_1y_2}}.
\eeq
It can be demonstrated that the parameter $A_y$ is directly proportional to the square root of the fugacity of ``species'' $y$~\cite{MS23b}.

    In Eqs.~\eqref{eq:eigen}--\eqref{eq:lambda} we have assumed a polydisperse system with a general mole fraction distribution $\phi^2_y$. On the other hand, contact with the original monocomponent q1D fluid necessitates the condition of equal chemical potential, i.e, $A_y=A$ for all $y$.
    In that case, Eq.~\eqref{eq:eigen} reduces to the eigenvalue/eigenfunction problem obtained from the transfer-matrix method~\cite{KP93}, where the (largest) eigenvalue $\ell$ is related to $A$ by $\ell = \beta p_\|/A^2$.
    Moreover, the excess Gibbs--Helmholtz free energy per particle of the equal-chemical-potential 1D polydisperse system becomes~\cite{MS23,MS23b}
    \begin{equation}\label{eq:gibbs}
	\beta \gi^\ex(\bp,\epsilon)=-\ln\frac{\ell(\bp,\epsilon)}{\epsilon}.
\end{equation}
Taking into account that $\lim_{\bp\to 0}\ell=\epsilon$ \cite{MS23}, we have that $\lim_{\bp\to 0}\beta \gi^\ex=0$, as it should be.

When tackling the numerical solution of the equations for the 1D polydisperse system, we considered $M$-component discrete mixtures. Specifically, within the discretized rendition of Eq.~\eqref{Fredh}, the evaluation of $G_{y_1y_2}(s)$ was directly achieved through matrix inversion. The results showed a linear correlation with $M^{-1}$, allowing for a subsequent extrapolation to $M \to \infty$  \cite{MS24}.

\paragraph*{Thermodynamic properties.}
Due to the pronounced anisotropy of the q1D fluid, the thermodynamic pressure becomes a tensor with two diagonal components ($P_\xx$ and $P_\yy$) along the longitudinal and transverse directions, respectively. Both components have dimensions of force per unit length, but each exhibits distinct behaviors. In the mapped 1D polydisperse system, only the 1D pressure, $p_\xx=\epsilon P_\xx$, possesses physical significance,  and $\epsilon$ simply represents the interval over which the ``species'' label runs. On the other hand,  upon reverting to the original q1D system, we can still utilize Eq.~\eqref{eq:gibbs} by interpreting $\gi^\ex(\bp,\epsilon)$ as the thermodynamic potential in a hybrid ensemble: isothermal-isobaric in the longitudinal direction and canonical in the transverse one. Consequently, the independent thermodynamic variables are the longitudinal pressure $P_\xx$ (or, equivalently, $p_\xx$) and the transverse length $\epsilon$, with their conjugate variables being the longitudinal length $L$ and the transverse pressure $P_\yy$, respectively. We can denote this ensemble with the set $\{N, p_\xx,L_\perp,T\}$. It is indeed noteworthy that the mapping from q1D to 1D systems not only yields the longitudinal properties of the original system but also its transverse ones.

The longitudinal compressibility factor, $Z_\xx\equiv \beta P_\xx L\epsilon/N=\bp/\lambda$, and the transverse compressibility factor, $Z_\yy\equiv \beta P_\yy L\epsilon/N=\beta P_\yy/(\lambda/\epsilon)$, can be obtained from the thermodynamic relations $Z_\xx=1+\bp ({\partial \beta\gi^\ex}/{\partial \bp})_\epsilon$ and $Z_\yy=1-\epsilon ({\partial \beta\gi^\ex}/{\partial \epsilon})_{\bp}$.
Starting from the mathematical identity $(\partial/\partial\epsilon)_{\bp}=(\partial/\partial\epsilon)_{\beta P_\|}-(\beta P_\|/\epsilon)(\partial/\partial\beta P_\|)_{\epsilon}$, we can straightforwardly derive Eq.~(9) of Ref.~\cite{VBG11}: $Z_\yy=Z_\xx-\epsilon ({\partial \beta\gi^\ex}/{\partial \epsilon})_{\beta P_\xx}$.
Based on the notation $\ii=\|,\perp$, the final results can be expressed as follows:
\beq
\label{eq:Zi}
	Z_\ii 	=1+\frac{\bp}{\ell}\int_{\epsilon}  d y_1 \int_{\epsilon}  d y_2\,\phi_{y_1}\phi_{y_2} \omega^\ii_{y_1y_2}e^{-\bp a_{y_1y_2}},
\eeq
with $\omega^{\xx}_{y_1y_2}=a_{y_1y_2}$ and  $\omega_{y_1y_2}^{\yy}=y_{12}^2/a_{y_1y_2}$.
Equation~\eqref{eq:Zi} with $\ii=\|$ coincides with Eq.~\eqref{eq:lambda} after setting $A_y=A=\sqrt{\bp/\ell}$ in the latter. Moreover, it can be proved that Eq.~\eqref{eq:Zi} with $\ii=\perp$ is equivalent to the  contact-theorem expression $Z_\yy=\epsilon \phi_{\epsilon/2}^2$~\cite{DFS23,FS24}.

\paragraph*{Low-pressure behavior.}
Virial expansions stand out as one of the most common approaches for characterizing fluids under low-density conditions. Obtaining the exact virial coefficients, particularly those of lower order, remains essential  to understand the behavior of the system, as well as to validate the precision of approximate methodologies.
In our q1D fluid, the virial coefficients for each component of the compressibility factor are traditionally defined based on the expansion in powers of density, i.e., $Z_\ii = 1+\sum^\infty_{k=2} B_{k\ii}\lambda^{k-1}$. However, for practical convenience, it is far more advantageous to employ coefficients ${B'_{k\ii}}$ in the expansion expressed in terms of the longitudinal pressure~\cite{M14b,MS23}, namely $Z_\ii = 1+\sum^\infty_{k=2} B'_{k\ii}(\bp)^{k-1}$. Both sets of coefficients are simply related: $B_{2\ii}=B_{2\ii}'$, $B_{3\ii}=B_{2\xx}B_{2\ii}+B_{3\ii}'$, $B_{4\ii}=B_{2\xx}^2B_{2\ii}+2B_{2\xx}B_{3\ii}'+B_{3\xx}'B_{2\ii}+B_{4\ii}'$, \ldots.
Coefficients ${B_{k\xx}'}$, with $k=2,3,4$, are already known~\cite{KMP04,MS23}. To obtain ${B_{k\yy}'}$, it is only necessary to take into account the thermodynamic relation $\bp(\partial Z_\yy/\partial\bp)_\epsilon=-\epsilon(\partial Z_\xx/\partial\epsilon)_{\bp}$, yielding $B_{k\yy}'=-(k-1)^{-1}\epsilon\partial B_{k\xx}'/\partial \epsilon$. The final results are
\begin{subequations}
\label{Bnxy}
\begin{equation}
	B_{2\xx}=\frac{2}{3}\frac{ \left(1+\frac{\epsilon ^2}{2}\right) \sqrt{1-\epsilon ^2}-1}{\epsilon ^2}+\frac{\sin ^{-1}(\epsilon )}{\epsilon },
\end{equation}
\beq
	B_{2\yy}=\frac{4}{3}\frac{ \left(1-\frac{\epsilon ^2}{4}\right) \sqrt{1-\epsilon ^2}-1}{\epsilon ^2}+\frac{\sin ^{-1}(\epsilon )}{\epsilon },
\eeq
	\beq
	\label{eq:B3pxy}
	B_{3\ii}'=3B_{2\xx}B_{2\ii}-2W_{2\ii}+C_{2\ii},
	\eeq
\bal
\label{eq:B4pxy}
B_{4\ii}'=&B_{2\ii}\left(10B_{2\xx}^2-4W_{2\xx}+\frac{1}{2}C_{2\xx}\right)+3W_{3\ii}\nn
&-B_{2\xx}\left(8W_{2\ii}-C_{2\ii}\right)+C_{3\ii},
\eal
\end{subequations}
where $C_{2\xx}\equiv \frac{\epsilon^2}{6}-1$,	$C_{2\yy}\equiv-\frac{\epsilon^2}{6}$, $C_{3\xx}\equiv\frac{1+5\epsilon^2-(1-\epsilon^2)^{5/2}}{15\epsilon^2}$, and $C_{3\yy}\equiv\frac{2-(1-\epsilon^2)^{3/2}(2+3\epsilon^2)}{45\epsilon^2}$ are exact coefficients, while
\begin{subequations}
	\beq
	\label{eq:W_2xy}
	W_{2\ii} \equiv \frac{1}{\epsilon}\int_\epsilon {d}y \, \psi_y^{\xx}\psi_y^{\ii},
	\eeq
	\beq
	\label{eq:W_3xy}
	W_{3\ii} \equiv\frac{1}{3\epsilon^2}\int_\epsilon {d}y_1\int_\epsilon {d}y_2 \, \psi_{y_1}^{\xx}\left(2a_{y_1y_2}\psi_{y_2}^{\ii}+\omega^{\ii}_{y_1y_2}\psi_{y_2}^{\xx}\right)
	\eeq
\end{subequations}
are numerical integrals, with $\psi_y^{\|,\perp}\equiv\frac{1}{2\epsilon}\left(\bar{\psi}_y^\pm+\bar{\psi}_{-y}^\pm\right)$ and $\bar{\psi}_y^\pm\equiv\sin^{-1}\left(\frac{\epsilon}{2}+ y\right)\pm\left(\frac{\epsilon}{2}+ y\right)\sqrt{1-\left(\frac{\epsilon}{2}+ y\right)^2}$.

\begin{figure}
	\includegraphics[width=\columnwidth]{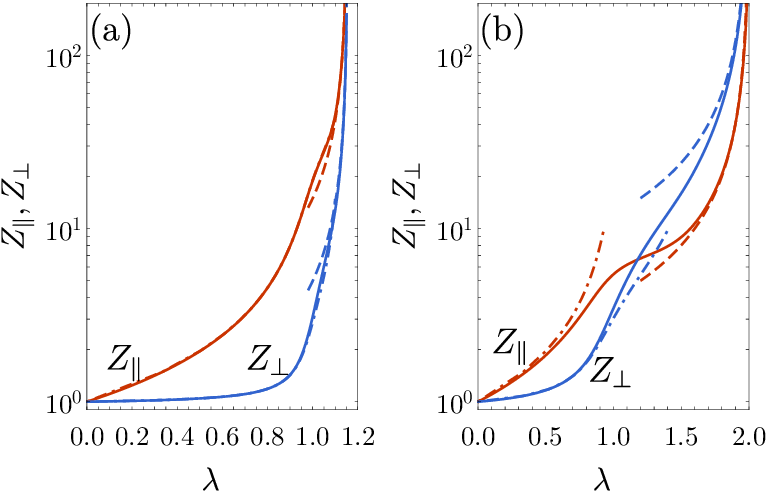}
	\caption{Plot of $Z_\xx$ and $Z_\yy$ as functions of the linear density for (a) $\epsilon=0.5$ and (b) $\epsilon=\sqrt{3}/2$. Dash dotted lines represent the truncated expansions $Z_\ii=1+\sum_{k=2}^4 B_{k\ii}'(\bp)^{k-1}$, while dashed lines represent the high-pressure behavior given by Eq.~\eqref{eq:limZi}.
We have checked that the figure is fully consistent with Fig.~3 of Ref.~\cite{VBG11}.}
	\label{fig:LimZ}
\end{figure}

\paragraph*{High-pressure behavior.}
In the limit $\beta p_\xx\to\infty$,  the linear density tends to its close-packing value $\lambda_{\text{cp}}=(1-\epsilon^2)^{-1/2}$. The corresponding asymptotic behaviors of $\phi_y$ and $\ell$ in that limit were derived in Ref.~\cite{MS23}.
Application of the limit in Eq.~\eqref{eq:Zi} yields
\begin{align}\label{eq:limZi}
Z_\ii\to  \frac{2\mathcal{A}_\ii}{1-\lambda/\lambda_{\text{cp}}},\quad \mathcal{A}_\xx=1,\quad \mathcal{A}_\yy=\lambda_{\text{cp}}^2-1.
\end{align}
When examining the behaviors of the compressibility factor's components under both low and high densities, a notable observation emerges: while $Z_\yy < Z_\xx$ consistently holds in the low-density range, this relation becomes true in the high-density regime only if $\lambda_{\mathrm{cp}} < \sqrt{2}$. Consequently, when $\epsilon > 1/\sqrt{2}\simeq 0.707$, at least one crossing point between both components arises. This crossing is unique, as depicted in Fig.~\ref{fig:LimZ}, while lower values of the width parameter $\epsilon$ exhibit no such crossing.
To better understand this point, let us consider the common tangent of two disks that are in contact at
close packing, and define the angle $\theta=\cos^{-1}\epsilon$ that the common tangent makes with the walls.
If $\epsilon>1/\sqrt{2}$, then $\theta<45^\circ$, which explains why $P_\yy>P_\xx$ as $\lambda\to\lcp$, while the opposite happens if $\theta>45^\circ$.

Figure~\ref{fig:LimZ} additionally demonstrates that both the low- and high-pressure approximations exhibit excellent performance across a broad spectrum of densities, extending beyond just the limiting scenarios. However, it is worth noting that the validity range decreases as the channel width parameter, $\epsilon$, grows.

\paragraph*{Behavior under maximum confinement.}
At a fixed linear density $\lambda$, the excess pore width $\epsilon$ can be made arbitrarily small {only} if $\lambda\leq 1$.
Assuming $\lambda<1$ and considering $\epsilon\ll 1$ in the eigenvalue equation for $\phi_y$ and $\ell$, one derives $\phi_y\to\epsilon^{-1/2}\left[1+\frac{\beta p_\xx}{2}\left({y}^2-\frac{\epsilon^2}{12}\right)\right]$ and
$\ell\to e^{-\beta p_\xx}\epsilon\left(1+\beta p_\xx\frac{\epsilon^2}{12}\right)$. Substituting these expressions into Eq.~\eqref{eq:Zi}, we obtain
\beq
\label{eq:epsilonto0}
Z_\xx\to 1+\bp\left(1-\frac{\epsilon^2}{12}\right),\quad Z_\yy\to 1+\bp\frac{\epsilon^2}{6},\quad (\lambda<1),
\eeq
implying $Z_\xx\to (1-\lambda)^{-1}$ and $Z_\yy\to 1$ in the limit $\epsilon\to 0$ if $\lambda<1$.
These results for $\lambda<1$ agree with those recently obtained by Franosch and Schilling through a different approach~\cite{FS24}.

If, on the other hand, $\lambda>1$, the smallest possible value of $\epsilon$ is $\sqrt{1-\lambda^{-2}}$. As one approaches this minimum value, we can use Eq.~\eqref{eq:limZi} to obtain
\beq
Z_\ii\to\frac{2\mathcal{A}'_\ii}{\lambda}\left(\epsilon-\sqrt{1-\lambda^{-2}}\right)^{-1},\quad (\lambda>1),
\eeq
with $\mathcal{A}'_\xx={1}/{\sqrt{\lambda^2-1}}$, $\mathcal{A}'_\yy=\sqrt{\lambda^2-1}$. The borderline case $\lambda=1$ necessitates special consideration. In this scenario, after some algebra, one finds
\beq
Z_\xx\sim\epsilon^{-2},\quad Z_\yy\to 3,\quad (\lambda=1).
\eeq

\paragraph*{Pair distribution functions.}
In liquid-state theory, the radial distribution function (RDF) stands as a pivotal structural characteristic, elucidating the variation of local density concerning distance from a reference particle. However, in confined liquids, defining a global RDF, $g(r)$, proves less straightforward compared to bulk systems due to the loss of rotational invariance in the fluid.
In general, if $n_1(\mathbf{r})$ is the local number density and $n_2(\mathbf{r}_1,\mathbf{r}_2)$ is the two-body configurational distribution function, the pair correlation function $g(\mathbf{r}_1,\mathbf{r}_2)$ is defined by $n_2(\mathbf{r}_1,\mathbf{r}_2)=n_1(\mathbf{r}_1)n_1(\mathbf{r}_2)g(\mathbf{r}_1,\mathbf{r}_2)$. In the q1D fluid, $n_1(\mathbf{r})=\lambda\phi^2_y$ and $g(\mathbf{r}_1,\mathbf{r}_2)=g_{y_1,y_2}(x_{12})$, where $x_{12}=|x_1-x_2|$. The function $g_{y_1,y_2}(x)$ can be identified with the interspecies RDF  of the 1D polydisperse system, which, in Laplace space, is given by Eq.~\eqref{Fredh} with $A_y=\sqrt{\bp/\ell}$. The transverse-averaged longitudinal correlation function is expressed as $g_\xx(x)=\int_\epsilon d y_1\int_\epsilon d y_2\, \phi_{y_1}^2\phi_{y_2}^2g_{y_1,y_2}(x)$.

As an alternative to Eq.~\eqref{eq:Zi}, it is feasible to express the compressibility factors in terms of $\lambda$ and integrals involving $g_\xx(x)$. Specifically, $Z_\xx=(1-I_0)/[1-\lambda(1-I_0+I_1^+)]$ and
$Z_\yy=Z_\xx\left[\lambda(1-I_0+I_1^-)-1\right]+2+I_2^-$, where $I_n^\pm\equiv \lambda\int_{\sqrt{1-\epsilon^2}}^1  d x\, x^{\pm n}g_\xx(x)$.

Let us now define the radial pair distribution function, $\hat{n}(r)$, such $\hat{n}(r) d r$ is the average number of  particles at a distance between  $r$ and  $r+ d r$ from any other particle. As a marginal distribution, $\hat{n}$ is obtained from $n_2$ as $\hat{n}(r)=N^{-1}\int  d\mathbf{r}_1\int  d\mathbf{r}_2\, n_2(\mathbf{r}_1,\mathbf{r}_2) \delta\left(r-\sqrt{x_{12}^2+y_{12}^2}\right)$. After some algebra, and assuming $r\ll L$, one finds
\beq
\label{eq:n2b}
\hat{n}(r)=2\lambda r\int_{\epsilon}^\dagger d y_1 \int_{\epsilon}^\dagger d y_2\, \phi^2_{y_1}\phi^2_{y_2} \frac{g_{y_1y_2}\left(\sqrt{r^2-y_{12}^2}\right)}{  \sqrt{r^2-y_{12}^2}}  ,
\eeq
where the dagger symbolizes the constraint $y_{12}^2<r^2$ imposed on the integrals. In the regime $1\ll r\ll L$, where correlations are negligible,  there exist two stripes of height $\epsilon$ and width $ d r$ at a distance $r$ from a certain reference particle. As a consequence, $\hat{n}(r)\approx 2\lambda$ in that regime.
In an ideal gas, the absence of interactions yields $\phi^2_y \to \epsilon^{-1}$ and $g_{y_1y_2}(x)\to 1$, resulting in
\beq
	\hat{n}^\mathrm{id}(r)=\frac{4\lambda r}{\epsilon}
	\begin{cases}
		\frac{\pi}{2}-\frac{r}{\epsilon},&r\leq \epsilon,\\
		\sqrt{\left(\frac{r}{\epsilon}\right)^2-1}-\frac{r}{\epsilon}+\sin^{-1}\left(\frac{\epsilon}{r}\right),& r\geq \epsilon.
	\end{cases}
\eeq
Interestingly, $\hat{n}^\mathrm{id}(r)$ is not constant due to the pronounced anisotropy of the system. Now we return to the interacting fluid.  Neglecting spatial correlations (but retaining the actual density profile $\phi_y^2$) would yield $\hat{n}^{\mathrm{nc}}(r)$ by setting $g_{y_1y_2}(x)\to 1$  in Eq.~\eqref{eq:n2b}. The RDF of the q1D fluid can be defined as the ratio $g(r)=\hat{n}(r)/\hat{n}^{\mathrm{nc}}(r)$, which differs from the average function $\overline{g}(r)=\int_\epsilon  d y_1 \int_\epsilon  d y_2\, \phi^2_{y_1}\phi^2_{y_2}g_{y_1y_2}(\sqrt{r^2-y_{12}^2})$.

\begin{figure}
	\includegraphics[width=\columnwidth]{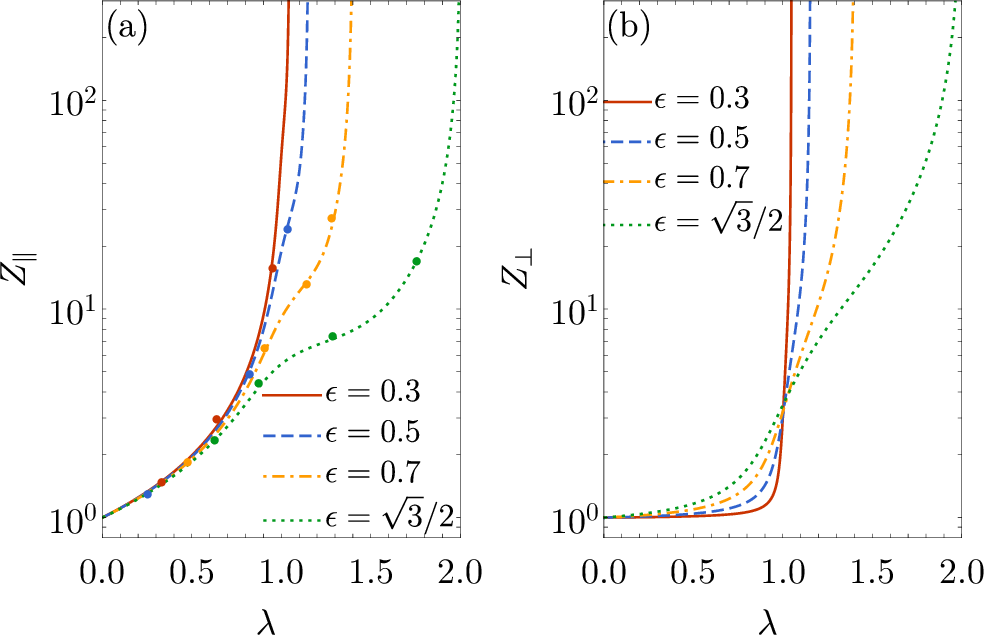}
	\caption{Plot of (a) $Z_\xx$  and (b) $Z_\yy$ as functions of the linear density for different values of the excess pore width $\epsilon$. The symbols in panel (a) represent data for $Z_\xx$ obtained  from $\{N,p_\xx,L_\yy,T\}$ MC simulations.}
	\label{fig:Z}
\end{figure}

\begin{figure}
	\includegraphics[width=\columnwidth]{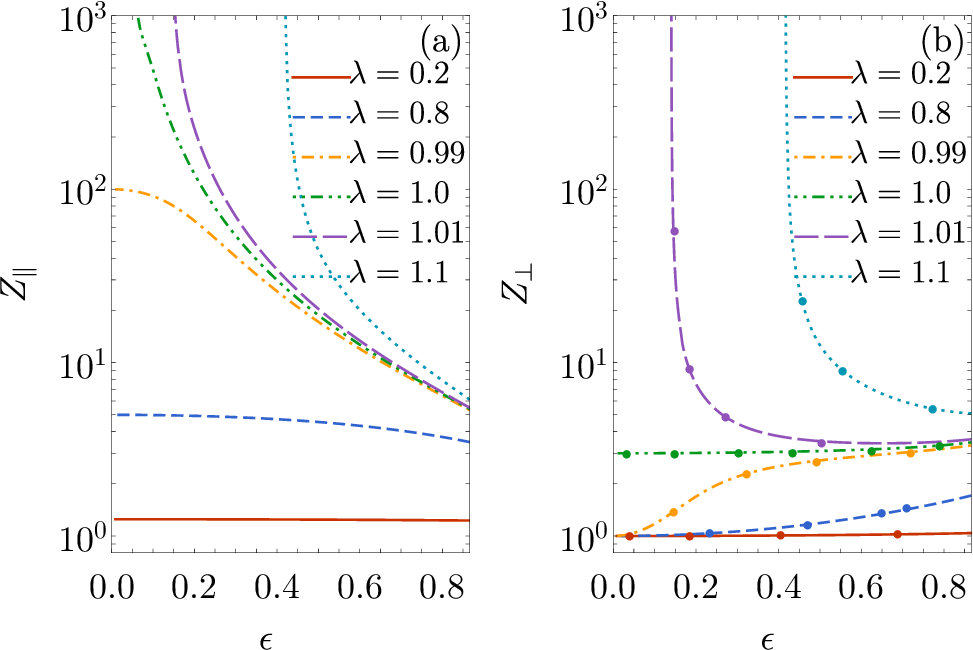}
	\caption{Plot of (a) $Z_\xx$  and (b) $Z_\yy$  as functions of the excess pore width  for different values of the linear density $\lambda$. The symbols in panel (b) represent data for $Z_\yy$ obtained  from $\{N,P_\yy,L_\xx,T\}$ MC simulations.}
	\label{fig:SimEps}
\end{figure}

\paragraph*{Validating theory through simulations.}
To validate the theoretical predictions derived within the 1D framework, Monte Carlo (MC) simulations were conducted on the original q1D fluid. For obtaining the longitudinal compressibility factor $Z_\xx$, simulations were performed in the $\{N,p_\xx,L_\yy,T\}$ ensemble, while the $\{N,P_\yy,L_\xx,T\}$ ensemble was utilized for determining $Z_\yy$. Conversely, the spatial correlation functions were assessed within the canonical $\{N,L_\xx,L_\yy,T\}$ ensemble. In general, $10^2$ particles were used and $10^7$ samples were collected after a sufficiently large equilibration process.

Figure~\ref{fig:Z} illustrates the density-dependence of the compressibility factors for various width parameter values. Both quantities exhibit divergence at the close-packing density $\lambda_{\mathrm{cp}}=(1-\epsilon^2)^{-1/2}$. Remarkably, there is an excellent agreement between the theoretical $Z_\xx$ and its corresponding MC values obtained in the $\{N,p_\xx,L_\yy,T\}$ ensemble. The latter ensemble is not appropriate to measure the transverse compressibility factor in simulations. Thus, Fig.~\ref{fig:Z} is complemented by  Fig.~\ref{fig:SimEps}, where the $\epsilon$-dependence of $Z_\xx$ and $Z_\yy$ is shown for various densities. Again,  an excellent agreement between theoretical and MC values of $Z_\yy$ is observed. Figure~\ref{fig:SimEps} also shows that, as discussed before, $Z_\xx$ and $Z_\yy$ for $\lambda>1$ diverge as $\epsilon$ approaches its minimum value $\sqrt{1-\lambda^{-2}}$, while both compressibility factors reach finite values in the limit $\epsilon\to 0$ if $\lambda<1$. In the special case $\lambda=1$, $Z_\xx$ diverges in that limit but $Z_\yy\to 3$.
Interestingly, $Z_\yy\approx 3$ at $\lambda=1$ for practically {any} value of $\epsilon$, as Figs.~\ref{fig:Z}(b) and \ref{fig:SimEps}(b) show.

\begin{figure}
	\includegraphics[width=\columnwidth]{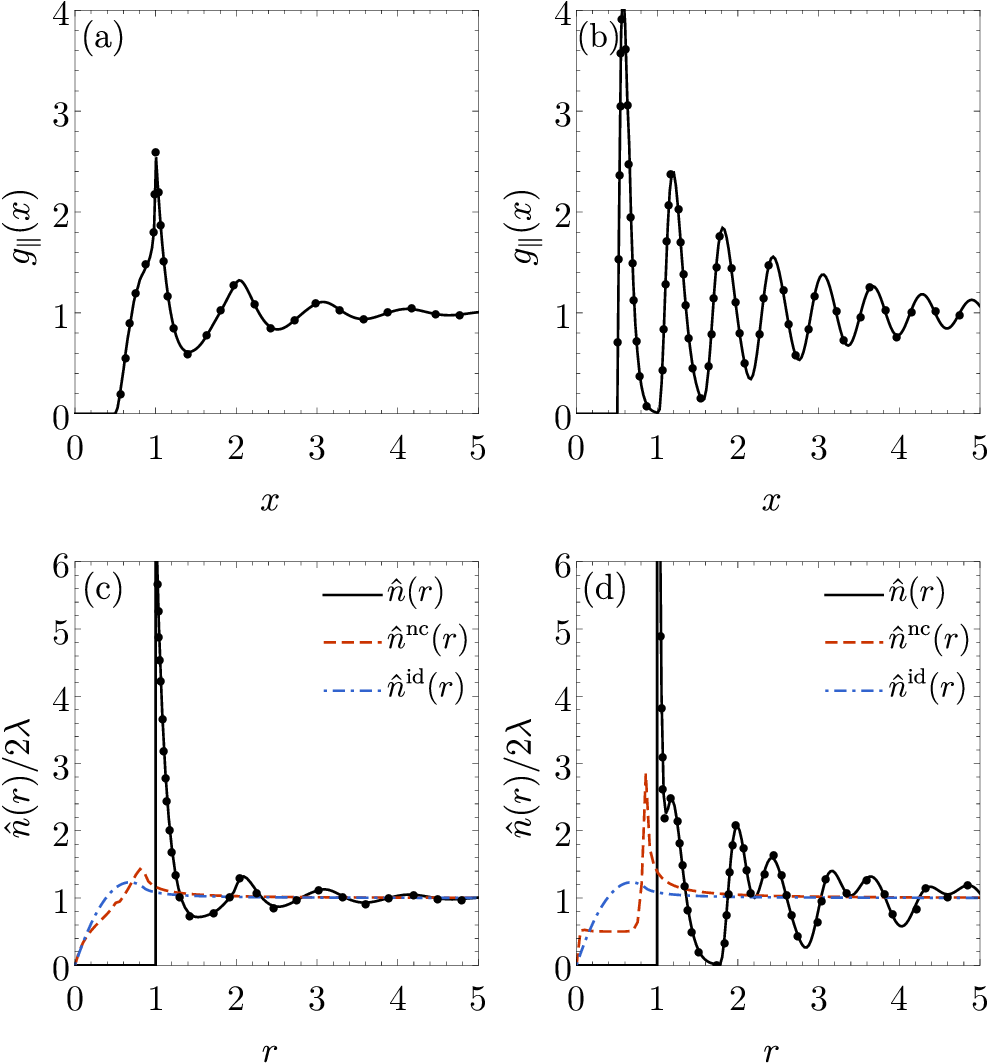}
	\caption{Plot of (a), (b) $g_\xx(x)$ and (c), (d) $\hat{n}(r)/2\lambda$ for  $\epsilon=\sqrt{3}/2$ and two density values: (a), (c) $\lambda=1.0$ and (b), (d) $\lambda=1.6$.  The symbols  represent data  obtained  from $\{N,L_\xx,L_\yy,T\}$ MC simulations. Panels (c) and (d) also include the functions $\hat{n}^{\text{nc}}(r)/2\lambda$ (dashed lines) and $\hat{n}^{\text{id}}(r)/2\lambda$ (dash dotted lines).}
	\label{fig:Gr}
\end{figure}

Now, let us examine the spatial correlation functions. Figure~\ref{fig:Gr} presents both the longitudinal correlation function, $g_\xx(x)$, and the radial pair distribution function, $\hat{n}(r)/2\lambda$, for $\epsilon=\sqrt{3}/2$ and two characteristic densities ($\lambda=1.0$ and $\lambda=1.6$). As expected, the MC simulations data  confirm the theoretical predictions for these correlation functions. It is evident that the structural characteristics of the q1D fluid exhibit considerably more complexity when transitioning from $\lambda=1.0$ to $\lambda=1.6$.
At $\lambda=1.6$, $g_\xx(x)$ displays evident oscillatory behavior, featuring local maxima positioned at
$x\simeq 0.58, 1.21, 1.81, 2.44, 3.07, 3.67, 4.30, 4.90,\ldots$,
consistent with the asymptotic wavelength of $0.63\simeq \lambda^{-1}$ derived from the dominant pole in Laplace space \cite{MS23b}.
Conversely, the oscillations of $\hat{n}(r)$ at $\lambda=1.6$ exhibit much less regularity, with local maxima  at $r=1$ and
$r\simeq 1.19, 1.99, 2.42, 3.17, 3.66, 4.38, 4.87,\ldots$. Significantly, the positions of the first, third, fifth, seventh, \ldots, maxima of $\hat{n}(r)$ and $g_\xx(x)$ are approximately related by the expression $r\simeq\sqrt{x^2+\epsilon^2}$. Conversely, the locations of the second, fourth, sixth, eighth, \ldots, maxima align with $r\simeq x$. These relations reveal a zigzag-like arrangement of the disks.
Figures~\ref{fig:Gr}(c) and \ref{fig:Gr}(d) additionally feature the ideal-gas radial function, $\hat{n}^{\text{id}}(r)/2\lambda$, and the one in the absence of correlations, $\hat{n}^{\text{nc}}(r)/2\lambda$. Both exhibit nonzero values and display a peak within the forbidden region $r<1$, swiftly approaching $1$ as $r>1$. Consequently, both ratios $\hat{n}(r)/\hat{n}^{\text{id}}(r)$ and $g(r)=\hat{n}(r)/\hat{n}^{\text{nc}}(r)$ are scarcely distinguishable from the plotted quantity $\hat{n}(r)/2\lambda$.

\paragraph*{Conclusions.}
Our investigation  delved into the nuanced properties of strongly confined hard-disk fluids within q1D channels, shedding light on both transverse and longitudinal behaviors. By leveraging an exact mapping onto a 1D polydisperse mixture of hard rods with equal chemical potentials, we unraveled various thermodynamic and structural characteristics across the whole spectrum of densities, thus providing a robust theoretical framework for our exploration. This equivalence, previously exploited only for longitudinal properties \cite{MS23,MS23b}, underscores the nontrivial nature of the confined system, characterized by a delicate balance between transverse confinement and inter-particle interactions. Supported by computer simulations, our findings offer valuable insights into the intricate properties of fluids in narrow channels, with implications for nanofluidics and experimental setups emulating such conditions. Moving forward, we hope that our work paves the way for further investigations into the transverse properties of such systems, bridging the gap between purely one-dimensional and bulk two- or three-dimensional systems. By elucidating the intricate interplay of confinement and interactions in q1D fluids, this work may contribute to the ongoing quest for a unified methodology to analyze and understand these complex systems.

\paragraph*{Acknowledgments.}
We acknowledge financial support from Grant No.~PID2020-112936GB-I00 funded by MCIN/AEI/10.13039/501100011033, and from Grant No.~IB20079  funded by Junta de Extremadura (Spain) and by European Regional Development Fund (ERDF) ``A way of making Europe.''
A.M.M. is grateful to the Spanish Ministerio de Ciencia e Innovaci\'on for a predoctoral fellowship Grant No.~PRE2021-097702.


\bibliography{C:/AA_D/Dropbox/Mis_Dropcumentos/bib_files/liquid}

\end{document}